\documentclass[aps,12pt,tightenlines,showpacs,showkeys]{revtex4}
\usepackage{graphicx}
\usepackage{dcolumn}
\begin{document}

\title{ Composite Photon Theory Versus Elementary Photon Theory }

\author{W. A. Perkins}

\affiliation {Perkins Advanced Computing Systems,\\ 12303 Hidden Meadows
Circle, Auburn, CA 95603, USA\\E-mail: wperkins@nccn.net} 

\begin{abstract}
The purpose of this paper is to show that the composite photon theory measures up well against the Standard Model's elementary photon theory. This is done by comparing the two theories area by area. Although the predictions of quantum electrodynamics are in excellent agreement with experiment (as in the anomalous magnetic moment of the electron), there are some problems, such as the difficulty in describing the electromagnetic field with the four-component vector potential because the photon has only two polarization states. In most areas the two theories give similar results, so it is impossible to rule out the composite photon theory. Pryce's arguments in 1938 against a composite photon theory are shown to be invalid or irrelevant. Recently, it has been realized that in the composite theory the antiphoton does not interact with matter because it is formed of a neutrino and an antineutrino with the wrong helicity. This leads to experimental tests that can determine which theory is correct.
\end{abstract}



\maketitle

\section{\label{sec.intro}Introduction}

In the history of physics many particles, which were once believed to be elementary, later turned out to be composites. The idea that the photon is a composite particle dates back to 1932, when Louis de Broglie~\cite{broglie1, broglie2} suggested that the photon is composed of a neutrino-antineutrino pair bound together. Pascual Jordan~\cite{jordan1}, who developed canonical anticommutation relations for fermions, thought he could obtain Bose commutation relations for a composite photon from the fermion anticommutation relations of its constituents. In order to obtain Bose commutation relations, Jordan modified de Broglie's theory, suggesting that a single neutrino could simulate a photon by a Raman effect and that no interaction between the neutrino and antineutrino was needed if they were emitted in exactly the same direction. Today, of course, we know that a single neutrino interacts much too weakly to simulate a photon. 
Because of Jordan's idea that the neutrino and antineutrino do not interact, the composite photon theory, was referred to as the ``Neutrino Theory of Light''.

Jordan's modifications made it easy for Pryce in 1938 to show that the theory was untenable. Pryce~\cite{pryce1} showed that if the composite photon obeyed Bose commutations relations, its amplitude would be zero. Pryce gave several arguments against the composite theory, but as Case~\cite{case1}, and Berezinskii~\cite{berezinskii1} discussed, the only valid argument was that the composite photon could not satisfy Bose commutation relations. In 1938 the existence of many other subatomic composite bosons, that are formed of fermion-antifermion pairs, was unknown. Perkins~\cite{perkins1} has shown that there is no need for a composite photon to satisfy exact Bose commutation relations. He points out that many composite bosons, such as Cooper pairs, deuterons, pions, and kaons, are not perfect bosons because of their internal fermion structure, although in the asymptotic limit they are essentially bosons. 

Neutrino oscillations in which one flavor of neutrino changes into another have been observed at the SuperKamiokande~\cite{fukuda} 
and SNO~\cite{ahmad}. Among the electron, muon, and tau neutrinos, at least two must have mass. Here we will assume that the composite photon is formed of an electron neutrino and an electron antineutrino and that the electron neutrinos are massless.

There has been some continuing work on the composite photon theory (see~\cite{dvoeglazov1,dvoeglazov2,perkins2}), but it still has not been accepted as an alternative to the elementary photon theory.
A major problem for the composite photon theory is that no experiment has demonstrated the need for it. Recently, Perkins~\cite{perkins4} showed that in the composite theory the antiphoton is different than the photon, and that antiphotons do not interact with electrons because their neutrinos have the wrong helicity.  This leads to experimental predictions that can differentiate between the Standard Model elementary photon theory and the composite photon theory. In the antihydrogen experiments at CERN the ALPHA~\cite{andresen1,amole1} and ASACUSA~\cite{enomoto1} Groups will be looking for spectral emissions from the antihydrogen atoms and shinning light on the atoms to put them in excited states. According to the composite photon theory, neither of these experiments will produce the expected results. 

In the next section we will compare the elementary and composite theories, area by area. In Section~\ref{sec.conclusions} we re-examine Pryce's arguments~\cite{pryce1} that the ``Neutrino Theory of Light'' is untenable and confirm that his arguments are no longer valid.

\section{\label{sec.compare} Comparison of Photon Theories}

Intuitively, de Broglie's idea makes reasonable the significant difference in characteristics exhibited by spin-1 photon and a spin-1/2 neutrino. When a photon is emitted, a neutrino-antineutrino pair arises from the vacuum. Later the neutrino and antineutrino annihilate when the photon is absorbed.

In the following sections we will examine the similarities and differences of the elementary and composite photon theories. Although the composite and elementary theories are similar, there are both subtle and major differences.

\subsection{Photon Field}

\subsubsection{ \label{ephotonfield} Elementary Photon Theory }

In noting the problem of quantizing the electromagnetic field, Bjorken and Drell~\cite{bjorken1} declared, ``It is ironic that of the fields we shall consider it is the most difficult to quantize.'' Srednicki~\cite{srednicki1} commented, ``Since real spin-1 particles transform in the $(1/2,1/2)$ representation of the Lorentz group, they are more naturally described as bispinors $A_{\alpha {\dot \alpha}}$ than as 4-vectors $A_\mu(x)$.'' Varlamov~\cite{varlamov} also noted that, ``the electromagnetic four-potential is transformed within $(1/2,1/2)$ representation of the homogeneous Lorentz group...'' Usually a canonical procedure for quantization is used although it is not manifestly covariant. We can describe the electromagnetic field with the four-component vector potential, but the photon only has two polarization states. One method of handling the problem is to introduce two non-physical photons along with the real ones, the 
Gupta-Bleuler procedure~\cite{schweber1}. Another method is to give the photon a very, very small mass~\cite{veltman1}. Following Bjorken and Drell~\cite{bjorken1} we will take only the transverse components and ``abandon manifest covariance.'' We start with Maxwell equations (in the absence of source charges and currents),
\begin{eqnarray}
\nabla \cdot {\bf E}(x) = 0, \nonumber \\
\nabla \cdot {\bf H}(x) = 0, \nonumber \\
\nabla \times {\bf E}(x) = - \partial {\bf H}(x) / \partial t, \nonumber \\
\nabla \times {\bf H}(x) = \partial {\bf E}(x) / \partial t. 
\label{eqn13ab}
\end{eqnarray}
This implies a vector potential, $A_{\mu} = ({\bf A},\phi)$, that satisfies,
\begin{eqnarray}
{\bf E}(x) = - { \partial {\bf A}(x) / \partial t } - \nabla \phi, \nonumber \\
{\bf H}(x) = \nabla \times {\bf A}(x).
\end{eqnarray}
For any electromagnetic field, $E$ and $H$, there are many $A_{\mu}$'s that differ by a gauge transformation.

A satisfactory Lagrangian density is given by,
\begin{eqnarray}
{\cal L}= -{1 \over 2} \left( {\partial A_\mu \over \partial x_\nu}
- {\partial A_\nu \over \partial x_\mu} \right)
 \left( {\partial A_\mu \over \partial x_\nu} \right).
\label{eqn13abc}
\end{eqnarray}
Using the standard method, we construct conjugate momenta 
from ${\cal L}$,
\begin{eqnarray}
\pi_0 = { \partial {\cal L} \over \partial { \dot{A}_0 }} = 0, \nonumber\\ 
\pi_k = { \partial {\cal L} \over \partial { \dot{A}_k }} 
= -\dot{A}_k -{\partial A_0 \over \partial x_k } = E_k.
\end{eqnarray}

\subsubsection{ Composite Photon Theory }

We start with the neutrino field. Solving the Dirac equation for a massless particle, 
$\gamma_\mu p_\mu \Psi = 0$, with $\Psi = u({\bf p})e^{ipx}$, results in the spinors,
\begin{eqnarray}
u^{+1}_{+1}({\bf p}) = \sqrt{ {E + p_3} \over 2 E} 
\left( \begin{array}{c}
1 \\ {{p_1 + i p_2} \over {E + p_3}} \\
0 \\ 0
\end{array} \right), 
\nonumber \\
u^{-1}_{-1}({\bf p}) = \sqrt{ {E + p_3} \over 2 E} 
\left( \begin{array}{c}
{{-p_1 + i p_2} \over {E + p_3}} \\ 1 \\
0 \\ 0 
\end{array} \right), 
\nonumber \\
u^{-1}_{+1}({\bf p}) = \sqrt{ {E + p_3} \over 2 E} 
\left( \begin{array}{c}
0 \\ 0 \\
1 \\ {{p_1 + i p_2} \over {E + p_3}} \\
\end{array} \right),
\nonumber \\
u^{+1}_{-1}({\bf p}) = \sqrt{ {E + p_3} \over 2 E} 
\left( \begin{array}{c}
0 \\ 0 \\  {{-p_1 + i p_2} \over {E + p_3}} \\ 1
\end{array} \right), 
\label{eqn6}
\end{eqnarray}
where $p_\mu = ({\bf p},iE)$, and the superscripts and subscripts on $u$ refer to the energy and helicity states respectively.
The gamma matrices in the Weyl basis were used in solving the Dirac equation:
\begin{eqnarray}
\gamma_1 = \left( \begin{array}{cccc}
0 & 0 & 0 & i \\
0 & 0 & i & 0 \\ 
0 & -i & 0 & 0 \\
-i & 0 & 0 & 0
\end{array} \right),
\; \; \; \; \gamma_2 = \left( \begin{array}{cccc}
0 & 0 & 0 & 1 \\
0 & 0 & -1 & 0 \\ 
0 & -1 & 0 & 0 \\
1 & 0 & 0 & 0
\nonumber \\
\end{array} \right),
\label{eqn2}
\end{eqnarray}
\begin{eqnarray}
\gamma_3 = \left( \begin{array}{cccc}
0 & 0 & i & 0 \\
0 & 0 & 0 & -i \\ 
-i & 0 & 0 & 0 \\
0 & i & 0 & 0
\end{array} \right),
\; \; \; \; \gamma_4 = \left( \begin{array}{cccc}
0 & 0 & 1 & 0 \\
0 & 0 & 0 & 1 \\ 
1 & 0 & 0 & 0 \\
0 & 1 & 0 & 0
\end{array} \right),
\label{eqn3}
\end{eqnarray}
\begin{eqnarray}
\gamma^5 = -\gamma_1 \gamma_2 \gamma_3 \gamma_4
= \left( \begin{array}{cccc}
1 & 0 & 0 & 0 \\
0 & 1 & 0 & 0 \\ 
0 & 0 & -1 & 0 \\
0 & 0 & 0 & -1
\end{array} \right).
\label{eqn4}
\end{eqnarray}
We designate $a_1$ as the annihilation operator for $\nu_1$, the right-handed neutrino, and $c_1$ as the annihilation operator for $\overline{\nu}_1$, the left-handed antineutrino. We assign $a_2$ as the annihilation operator for $\nu_2$, the left-handed neutrino, and $c_2$ as the annihilation operator for $\overline{\nu}_2$, the right-handed antineutrino. Since only $\nu_2$ and $\overline{\nu}_2$ have been observed, we take the neutrino field to be,
\begin{eqnarray}
\Psi(x) = {1 \over \sqrt{V}} \sum_{\bf k} \left\{ 
\left[ a_2({\bf k}) u^{+1}_{-1}({\bf k}) 
\right] e^{i k x} \right. 
\left. + \left[ 
c_2^\dagger({\bf k}) u^{-1}_{+1}({\bf -k}) \right]e^{-i k x} \right\},
\label{eqn5}
\end{eqnarray}
where we have used only the two corresponding spinors, and $k x$ stands for ${\bf k} \cdot {\bf x} - \omega_k t$. A four-vector field can be created from a fermion-antifermion pair,
\begin{equation}     
{\overline \Psi} i\gamma_{\mu} \Psi. 
\label{eqn7}
\end{equation}
The fermion and antifermion are bound by this attractive local vector interaction of Equation~(\ref{eqn7}) as discussed by Fermi and Yang~\cite{fermi-yang}. We postulate that this local interaction between the neutrino and antineutrino is responsible for their interaction with the electromagnetic coupling constant ``$\alpha$'' while a single neutrino interacts with the weak coupling constant ``g''. Both Kronig~\cite{kronig1} and de Broglie~\cite{broglie1} suggested local interactions in their work on the composite photon theory.
Since the neutrino and antineutrino momenta are in opposite directions, we take the photon field to be~\cite{perkins2}, 
\begin{eqnarray}
A_\mu(x) =  \sum_{\bf p} {-1 \over 2\sqrt{ V \omega_p}}\left\{ 
\left[G_R({\bf p}) {\overline u}^{+1}_{-1}({\bf p}) i \gamma_{\mu} 
u^{-1}_{+1}({\bf p})
+ G_L({\bf p}) {\overline u}^{-1}_{+1}({\bf p}) i \gamma_{\mu} 
u^{+1}_{-1}({\bf p}) 
\right]e^{i p x} \right. \nonumber \\
\left. + \left[G_R^\dagger({\bf p}) 
{\overline u}^{-1}_{+1}({\bf p}) i \gamma_{\mu} u^{+1}_{-1}({\bf p})
+ G_L^\dagger({\bf p}) {\overline u}^{+1}_{-1}({\bf p}) i \gamma_{\mu} 
u^{-1}_{+1}({\bf p}) 
\right]e^{-i p x}  \right\},
\label{eqn8}
\end{eqnarray}
with the annihilation operators for left-circularly and right-circularly polarized photons with momentum ${\bf p}$ given by,
\begin{eqnarray}
G_L( {\bf p}) = {1 \over \sqrt{2}} \sum_{\bf k} F^\dagger( {\bf k}) 
c_2( {\bf -k}) a_2( {\bf p} + {\bf k})  \nonumber \\
G_R( {\bf p}) = {1 \over \sqrt{2}} \sum_{\bf k} F^\dagger( {\bf k}) 
c_2( {\bf p} + {\bf k}) a_2( {\bf -k}),   
\end{eqnarray}
where $F({\bf k})$ is a spectral function. 

Although many sets of gamma matrices satisfy the Dirac equation, one must use the Weyl representation of gamma matrices to obtain spinors appropriate for the composite photon. If a different set of gamma matrices is used, the photon field will NOT satisfy Maxwell equations. Kronig\cite{kronig1} was the first to realize this, but he did not mention the deeper significance, i.e., two-component neutrinos are required for a composite photon. At that time a two-component neutrino theory would have been rejected because it violated parity. The connection between the photon antisymmetric tensor and the two-component Weyl equation was also noted by Sen~\cite{sen1}.
 Although we are working at the four-component level, one can form a composite photon at the two-component level~\cite{perkins2}.

\subsection{\label{elem_com_rel}Commutation Relations}

\subsubsection{ Elementary Photon Theory }

In classical Hamiltonian mechanics, the Poisson bracket is defined as,
\begin{eqnarray}
\left[F,G\right]_{PB} = {\partial F \over \partial q_k}{\partial G\over \partial p_k}
-{\partial F \over \partial p_k}{\partial G\over \partial q_k},
\label{eqn16ab}
\end{eqnarray}
where $q_k(t)$ are the generalized coordinate and $p_k(t)$ are the generalized momenta. If we use $q_i$ and $p_j$ in place $F$ and $G$,
we obtain the fundamental Poisson brackets,
\begin{eqnarray}
\left[ q_i(t),q_j(t) \right]_{PB} = 0, \nonumber \\
\left[p_i(t),p_j(t)\right]_{PB} = 0, \nonumber \\
\left[q_i(t),p_j(t)\right]_{PB} = \delta_{ij}.
\label{eqn16bc}
\end{eqnarray}
In going over to quantum theory, it is hypothesized that the fundamental Poisson brackets become commutators with $q_i$, and $p_i$ becoming operators,
\begin{eqnarray}
\left[q_i(t),q_j(t)\right] = 0, \nonumber \\
\left[p_i(t),p_j(t)\right] = 0, \nonumber \\
\left[q_i(t),p_j(t)\right] = i\delta_{ij}.
\label{eqn16bd}
\end{eqnarray}
The generalized coordinates and momenta for the classical electromagnetic field are,
\begin{eqnarray}
q_i(t) \rightarrow A_{\mu}({\bf x},t), \nonumber \\
p_j(t) \rightarrow \pi_{\mu}({\bf x},t). 
\label{eqn16bcd}
\end{eqnarray}
Thus, the fundamental commutators become,
\begin{eqnarray}
\left[A_{\mu}({\bf x},t),A_{\nu}({\bf x{'}},t)\right] = 0, \nonumber \\
\left[\pi_{\mu}({\bf x},t),\pi_{\nu}({\bf x{'}},t)\right] = 0, \nonumber \\
\left[\pi_{\mu}({\bf x},t),A_{\nu}({\bf x{'}},t)\right] 
=  -i\delta_{\mu\nu}\delta^3({\bf x}-{\bf x{'}}).
\label{eqn16cde}
\end{eqnarray}
However, the third Equation of~(\ref{eqn16cde}) is not consistent with Maxwell equations, so we must depart from the canonical path~\cite{bjorken1} and replace it with,
\begin{eqnarray}
\left[\pi_{\mu}({\bf x},t),A_{\nu}({\bf x{'}},t)\right] 
=  +i\delta^{tr}_{\mu\nu}({\bf x}-{\bf x{'}}).
\label{eqn16def}
\end{eqnarray}
Expanding the ${\bf A}$ and ${\bf \pi}$ into plane waves,
\begin{eqnarray}
{\bf A}(x) =  \int { {d^3p  \over \sqrt{ 2 \omega_p (2 \pi)^3)}} 
\sum_{\lambda = 1}^2 
{\bf \epsilon}^{\lambda}({\bf p})
\left[ b_{\lambda}({\bf p}) e^{-i p x}
+ b^\dagger_{\lambda}({\bf p})e^{i p x} \right]}, \nonumber\\
{\bf \pi}(x) = {\bf {\dot A}} =  i\int { {d^3p  
\sqrt{ {\omega_p \over 2(2 \pi)^3)}}} \sum_{\lambda = 1}^2 
{\bf \epsilon}^{\lambda}({\bf p})
\left[ -b_{\lambda}({\bf p}) e^{-i p x}
+ b^\dagger_{\lambda}({\bf p})e^{i p x} \right]}.
\label{eqn8.4ab}
\end{eqnarray}
where $\omega_p = p_0$, and $b_{\lambda}({\bf p})$ and
$b^\dagger_{\lambda}({\bf p})$ are identified as annihilation and creation operators for polarization $\lambda$.  We take the two unit polarization vectors to be perpendicular to ${\bf p}$ in order to satisfy
$\nabla \cdot {\bf A}(x) = 0$ (i.e., radiation gauge),
\begin{eqnarray}
{\bf \epsilon}^{\lambda}({\bf p}) \cdot {\bf p} = 0.
\label{eqn7.1ab}
\end{eqnarray}
Also it is convenient to choose,
\begin{eqnarray}
{\bf \epsilon}^{\lambda}({\bf p}) \cdot
{\bf \epsilon}^{\lambda^{'}}({\bf p})  = \delta_{\lambda \lambda^{'}}.
\label{eqn7.2ab}
\end{eqnarray}
Inverting Equation~(\ref{eqn8.4ab}) we obtain the amplitudes, $b_{\lambda}({\bf p})$ and $b^\dagger_{\lambda}({\bf p})$,
\begin{eqnarray}
b_{\lambda}({\bf p}) = \int  {{d^3x  \; e^{i p x}} \over 
{\sqrt(2 \omega_p (2 \pi)^3)}} {\bf \epsilon}^{\lambda}({\bf p}) 
\cdot \left[\omega_p {\bf A}(x) + i{\bf \dot{A}}(x)\right], \nonumber \\
b^\dagger_{\lambda}({\bf p}) = -\int  {{d^3x  \; e^{-i p x}} \over 
{\sqrt(2 \omega_p (2 \pi)^3)}} {\bf \epsilon}^{\lambda}({\bf p}) 
\cdot \left[\omega_p {\bf A}(x) + i{\bf \dot{A}}(x)\right].
\label{eqn8.4abc}
\end{eqnarray} 
Following Bjorken and Drell~\cite{bjorken1}, we use Equations~(\ref{eqn16cde}) and (\ref{eqn16def}) to obtain commutation relations for the annihilation and creation operators,
\begin{eqnarray}
\left[b_{\lambda}({\bf p}),b_{\lambda{'}}({\bf q})\right] = 0,  \nonumber \\
\left[b^\dagger_{\lambda}({\bf p}),b^\dagger_{\lambda{'}}({\bf q})\right] = 0,  \nonumber \\
\left[b_{\lambda}({\bf p}),b^\dagger_{\lambda{'}}({\bf q})\right] 
= \delta({\bf p}-{\bf q})\delta_{\lambda \lambda{'}}.
\label{eqn16}
\end{eqnarray}
Left-handed and right-handed circularly polarized annihilation operators are obtained from the combinations,
\begin{eqnarray}
b_L( {\bf p}) = {1 \over \sqrt{2}} \left[ b_1({\bf p})-i b_2({\bf p})\right], \nonumber\\
b_R( {\bf p}) = {1 \over \sqrt{2}} \left[ b_1({\bf p}) 
+i b_2({\bf p})\right], 
\end{eqnarray}
and they obey the commutation relations,
\begin{eqnarray}
\left[b_L({\bf p}),b_L({\bf q})\right] = 0, 
\left[b^\dagger_L({\bf p}),b^\dagger_L({\bf q})\right] = 0, \nonumber \\
\left[b_L({\bf p}),b^\dagger_L({\bf q})\right] 
= \delta({\bf p}-{\bf q}), \nonumber \\
\left[b_R({\bf p}),b_R({\bf q})\right] = 0, 
\left[b^\dagger_R({\bf p}),b^\dagger_R({\bf q})\right] = 0, \nonumber \\
\left[b_R({\bf p}),b^\dagger_R({\bf q})\right] 
= \delta({\bf p}-{\bf q}),  \nonumber \\
\left[b_L({\bf p}),b_R({\bf q})\right] = 0, 
\left[b_L({\bf p}),b^\dagger_R({\bf q})\right] = 0.
\label{eqn17zy}
\end{eqnarray}
From this discussion it is evident that the elementary photon commutation relations were carried over from the classical canonical formalism and are not based on any fundamental principle. The photon distribution for Blackbody radiation can be calculated using the second quantization method~\cite{koltun1}, including commutation relations of Equation~(\ref{eqn16}), resulting in Planck's law,
\begin{equation}
n_{\bf p} = {1 \over e^{\omega_p /kT} - 1},
\label{eqn16.5}
\end{equation}

\subsubsection{ Composite Photon Theory }

Composite integral spin particles obey commutation 
relations~\cite{sahlin1,lipkin1,perkins3} that are derived from the fermion anticommutation relations of their constituents.
For composite photons we have,
\begin{eqnarray}
\left[G_L({\bf p}),G_L({\bf q})\right] = 0, 
\left[G^\dagger_L({\bf p}),G^\dagger_L({\bf q})\right] = 0, \nonumber \\
\left[G_L({\bf p}),G^\dagger_L({\bf q})\right] 
= \delta({\bf p}-{\bf q})(1-\Delta_L({\bf p},{\bf p})),  \nonumber \\
\left[G_R({\bf p}),G_R({\bf q})\right] = 0, 
\left[G^\dagger_R({\bf p}),G^\dagger_R({\bf q})\right] = 0, \nonumber \\
\left[G_R({\bf p}),G^\dagger_R({\bf q})\right] 
= \delta({\bf p}-{\bf q})(1-\Delta_R({\bf p},{\bf p})),  \nonumber \\
\left[G_L({\bf p}),G_R({\bf q})\right] = 0, 
\left[G_L({\bf p}),G^\dagger_R({\bf q})\right] = 0, \nonumber \\
\Delta_L({\bf p},{\bf p}) = \sum_{\bf k} |F({\bf k})|^2 \left[  
a^\dagger_2({\bf p}+{\bf k})a_2({\bf p}+{\bf k}) \right. \nonumber \\
\left. + c^\dagger_2(-{\bf k})
c_2(-{\bf k}) \right], \nonumber \\
\Delta_R({\bf p},{\bf p}) = \sum_{\bf k} |F({\bf k})|^2 \left[  
a^\dagger_2(-{\bf k})a_2(-{\bf k}) \right. \nonumber \\
\left. + c^\dagger_2({\bf p}+{\bf k})
c_2({\bf p}+{\bf k}) \right]. 
\label{eqn17}
\end{eqnarray}
In obtaining the commutation relations involving $\Delta_R({\bf p},{\bf p})$ and
$\Delta_L({\bf p},{\bf p})$, we have taken the expectation values. 
Here the linearly-polarized photon annihilation operators are defined as,
\begin{eqnarray}
\xi( {\bf p}) = {1 \over \sqrt{2}} \left[ G_L({\bf p}) + G_R({\bf p})\right] \nonumber\\
\eta( {\bf p}) = {i \over \sqrt{2}} \left[ G_L({\bf p})-G_R({\bf p})\right] 
\end{eqnarray}
and they obey the commutation relations,
\begin{eqnarray}
\left[\xi({\bf p}),\xi({\bf q})\right] = 0, 
\left[\xi^\dagger({\bf p}),\xi^\dagger({\bf q})\right] = 0, 
\nonumber \\
\left[\xi({\bf p}),\xi^\dagger_({\bf q})\right] 
= \delta({\bf p}-{\bf q}) \left( 1-{1 \over 2} 
( \Delta_L({\bf p},{\bf p})
+ \Delta_R({\bf p},{\bf p})) \right), \nonumber \\
\left[\eta({\bf p}),\eta({\bf q})\right] = 0, 
\left[\eta^\dagger({\bf p}),\eta^\dagger({\bf q})\right] = 0, 
\nonumber \\
\left[\eta({\bf p}),\eta^\dagger({\bf q})\right] 
= \delta({\bf p}-{\bf q}) \left( 1-{1 \over 2} 
( \Delta_L({\bf p},{\bf p})
+ \Delta_R({\bf p},{\bf p})) \right), \nonumber \\
\left[\xi({\bf p}),\eta({\bf q})\right] = 0, \nonumber \\
\left[\xi({\bf p}),\eta^\dagger({\bf q})\right] 
= {i \over 2} \delta({\bf p}-{\bf q}) 
\left( \Delta_L({\bf p},{\bf p})-\Delta_R({\bf p},{\bf p}) \right). 
\label{eqn17ba}
\end{eqnarray}
One virtue of a good theory is simplicity. Although the composite photon commutations relations~(\ref{eqn17}) and~(\ref{eqn17ba}) appear more complex than the elementary commutations relations~(\ref{eqn16})
and~(\ref{eqn17zy}), they are really simpler because it is only necessary to postulate the fermion anticommutation relations and then derive boson commutation relations. A more detailed discussion is contained in Ref.~\cite{perkins1}.

The composite photon distribution for Blackbody radiation can be calculated using the second quantization method~\cite{koltun1} as above, but with the composite photon commutation relations. This results~\cite{perkins1} in, 
\begin{equation}
n_{\bf p} = {1 \over e^{\omega_p /kT}
\left( 1 + {1 \over \Omega} \right) - 1},
\label{eqn18.2}
\end{equation}
The $1 \over \Omega$ component is less than $10^{-9}$, so the difference between Equation~(\ref{eqn16.5}) and (\ref{eqn18.2}) is too small to measure.

\subsection{Polarization Vectors}
 In the elementary theory the polarization vectors are chosen so that the electromagnetic field satisfies Maxwell equations. In composite theory there is no flexibility; the polarization vectors are given by the neutrino bispinors. 

\subsubsection{ Elementary Photon Theory }

Polarization vectors for photons with spin parallel and antiparallel to their momentum (taken to be along the third axis) are given by,
\begin{eqnarray}     
\epsilon_\mu^1(n) = {1 \over \sqrt{2}}(1,i,0,0), \nonumber\\ 
\epsilon_\mu^2(n) = {1 \over \sqrt{2}}(1,-i,0,0).
\label{eqn10}
\end{eqnarray} 
In Section~\ref{elem_com_rel} we chose some properties of the polarization vectors in Equation~(\ref{eqn7.1ab}) 
and (\ref{eqn7.2ab}). In four dimensions we have,
\begin{eqnarray}     
\epsilon_\mu^j(p) \cdot \epsilon_\mu^{k*}(p) = \delta_{jk},
\label{eqn11a}
\end{eqnarray}     
and the dot products with the internal four-momentum $p_\mu$,
\begin{eqnarray}     
p_\mu \epsilon_\mu^1(p) = 0, \nonumber\\ 
p_\mu \epsilon_\mu^2(p) = 0.
\label{eqn12ab}
\end{eqnarray} 
Also in three dimensions,
\begin{eqnarray}     
{\bf p} \times {\bf \epsilon^1}({\bf p})=-i \omega_p 
{\bf \epsilon^1}({\bf p}), \nonumber\\ 
{\bf p} \times {\bf \epsilon^2}({\bf p})= i \omega_p 
{\bf \epsilon^2}({\bf p}).
\label{eqn27bc}
\end{eqnarray}     
To calculate the completeness relation, we use linear polarization vectors. Noting that the sum over polarization states only involves the two transverse polarizations and not the third direction ${\bf p}$,
\begin{equation}
\sum_{\lambda=1}^2 \epsilon_j^{\lambda}({\bf p}) \epsilon_l^{\lambda}({\bf p})
= \sum_{\lambda=1}^3 \epsilon_j^{\lambda}({\bf p})
\epsilon_l^{\lambda}({\bf p})-\epsilon_j^{3}({\bf p}) 
\epsilon_l^3({\bf p}) \nonumber \\
= \delta_{j l} - {p_j p_l \over p^2}. 
\label{eqn15cd}
\end{equation}

\subsubsection{\label{sec.composite_polar_vectors}Composite Photon Theory }

From Equation~(\ref{eqn8}) we see that the polarization vectors are neutrino bispinors:
\begin{eqnarray}
\epsilon_\mu^1( p ) = {-1 \over \sqrt{2}} [{\overline u}^{+1}_{-1}({\bf p})
i\gamma_{\mu} u^{-1}_{+1}({\bf p})],  \nonumber \\
\epsilon_\mu^2( p ) = {-1 \over \sqrt{2}} [{\overline u}^{-1}_{+1}({\bf p})  i\gamma_{\mu} u^{+1}_{-1}({\bf p})].
\label{eqn11}
\end{eqnarray}
Carrying out the matrix multiplications results in,
\begin{eqnarray}     
\epsilon_\mu^1(p) \!= \!{1 \over \sqrt{2}} \left( 
{{-i p_1 p_2 \!+\!E^2 \!+\!p_3 E \!-\!p_1^2} \over {E(E + p_3)}},
{{- p_1 p_2 \! + \!iE^2 \! +\!ip_3 E \! - \!ip_2^2 } 
\over {E(E + p_3)}},
{{\!-p_1 \!- \!i p_2} \over E}, 0 \right), \nonumber\\ 
\epsilon_\mu^2(p) \!= \!{1 \over \sqrt{2}} \left( 
{{i p_1 p_2 \!+\!E^2 \!+\!p_3 E \!-\!p_1^2} \over {E(E + p_3)}},
{{-p_1 p_2 \! - \!iE^2 \! -\!ip_3 E \! + \!ip_2^2 } 
\over {E(E + p_3)}},
{{\!-p_1 \!+ \!i p_2} \over E}, 0 \right), \nonumber\\ 
\label{eqn12}
\end{eqnarray}    
Since the neutrino spinors and the polarization vectors only depend upon the direction of 
${\bf p}$, we can set ${\bf n} = {\bf p}/ E$. 
\begin{eqnarray}     
\epsilon_\mu^1(n) \!= \!{1 \over \sqrt{2}} \left( 
{{-i n_1 n_2 \!+\!1 \!+\!n_3 \!-\!n_1^2} \over {1 + n_3}},
{{- n_1 n_2 \!+ \!in_1^2 \!+ \!in_3^2 \!+ \!in_3} 
\over {1 + n_3}},
\!-n_1 \!- \!i n_2, 0 \right), \nonumber\\ 
\epsilon_\mu^2(n) \!= \!{1 \over \sqrt{2}}\left( 
{{i n_1 n_2 \!+\!1 \!+\!n_3 \!-\!n_1^2} \over {1 + n_3}},
{{- n_1 n_2 \!- \!in_1^2 \!- \!in_3^2 \!- \!in_3} 
\over {1 + n_3}},
\!-n_1 \!+ \!i n_2, 0 \right).
\label{eqn12.5}
\end{eqnarray}     
As one can see these polarization vectors are good for any direction ${\bf n}$, while the elementary polarization vectors, Equation~(\ref{eqn10}), are only given along the third axis.
These polarization vectors satisfy the normalization relation,
\begin{eqnarray}
\epsilon_\mu^j(p) \cdot \epsilon_\mu^{k*}(p) = \delta_{jk},
\label{eqn11ax}
\end{eqnarray}     
and the dot products with the internal four-momentum $p_\mu$ give,
\begin{eqnarray}     
p_\mu \epsilon_\mu^1(p) = 0, \nonumber\\ 
p_\mu \epsilon_\mu^2(p) = 0.
\label{eqn12abx}
\end{eqnarray} 
Also in three dimensions,
\begin{eqnarray}     
{\bf p} \times {\bf \epsilon^1}({\bf p})=-i \omega_p 
{\bf \epsilon^1}({\bf p}), \nonumber\\ 
{\bf p} \times {\bf \epsilon^2}({\bf p})= i \omega_p 
{\bf \epsilon^2}({\bf p}).
\label{eqn27bcx}
\end{eqnarray}

Using Equation~(\ref{eqn12}) we calculate the completeness relation,
\begin{equation}
\sum_{j=1}^2 \epsilon_{\mu}^j({\bf p}) \epsilon_{\nu}^{j*}({\bf p})
= \sum_{j=1}^2 \epsilon_{\mu}^{j*}({\bf p}) \epsilon_{\nu}^j({\bf p})
= \delta_{\mu \nu} - {p_{\mu} p_{\nu} \over E^2}. 
\label{eqn15}
\end{equation}

\subsection{Maxwell Equations}

\subsubsection{ Elementary Photon Theory }

In the elementary theory, Maxwell equations are taken as an experimental result as discussed in Section~\ref{ephotonfield}. The vector potential, $A_\mu(x)$, is then created to satisfy Maxwell equations.

\subsubsection{ \label{sec.maxwell} Composite Photon Theory }

In the composite theory, Maxwell equations are derived, as they must be if the composite theory is relevant. 
Substituting Equation~(\ref{eqn11}) into Equation~(\ref{eqn8}) gives $A_\mu$ in terms of the polarization vectors, 
\begin{eqnarray}
A_\mu(x) =  \sum_{\bf p} {1 \over \sqrt{ 2 V \omega_p}}\left\{ 
\left[G_R({\bf p}) \epsilon_{\mu}^1({\bf p})
+ G_L({\bf p}) \epsilon_{\mu}^2({\bf p})
\right]e^{i p x} \right. \nonumber \\
\left. + \left[G_R^\dagger({\bf p}) \epsilon_{\mu}^{1*}({\bf p})
+ G_L^\dagger({\bf p}) \epsilon_{\mu}^{2*}({\bf p})
\right]e^{-i p x}  \right\}.
\label{eqn8.4}
\end{eqnarray}
The electric and magnetic fields are obtained from 
${\bf E}(x) = - { \partial {\bf A}(x) / \partial t }$ and
${\bf H}(x) = \nabla \times {\bf A}(x)$ as usual,
\begin{eqnarray}
E_\mu(x) =  i \sum_{\bf p} {\sqrt{\omega_p} \over \sqrt{2 V }}\left\{ 
\left[G_R({\bf p}) \epsilon_\mu^1({\bf p})
+ G_L({\bf p}) \epsilon_\mu^2({\bf p})  \right]e^{i p x} 
\right. \nonumber \\
\left. - \left[G_R^\dagger({\bf p}) \epsilon_\mu^{1*}({\bf p})
+ G_L^\dagger({\bf p}) \epsilon_\mu^{2*}({\bf p})  
\right]e^{-i p x}  \right\},
\label{eqn5abc}
\end{eqnarray}
\begin{eqnarray}
H_\mu(x) =  \sum_{\bf p} {\sqrt{\omega_p} \over \sqrt{2 V }}\left\{ 
\left[
G_R({\bf p}) \epsilon_\mu^1({\bf p})  
- G_L({\bf p}) \epsilon_\mu^2({\bf p})  \right]e^{i p x} 
\right. \nonumber \\
\left. + \left[G_R^\dagger({\bf p}) \epsilon_\mu^{1*}({\bf p})
- G_L^\dagger({\bf p}) \epsilon_\mu^{2*}({\bf p})  
\right]e^{-i p x}  \right\}.
\label{eqn5abcd}
\end{eqnarray}
Using Equation~(\ref{eqn12abx}) we obtain,
\begin{eqnarray}
\nabla \cdot {\bf E}(x) = 0, \nonumber \\
\nabla \cdot {\bf H}(x) = 0. 
\label{eqn13.5}
\end{eqnarray}
and with Equation~(\ref{eqn27bcx}) we obtain,
\begin{eqnarray}
\nabla \times {\bf E}(x) = - \partial {\bf H}(x) / \partial t, \nonumber \\
\nabla \times {\bf H}(x) = \partial {\bf E}(x) / \partial t. 
\label{eqn13}
\end{eqnarray}
Using Equation~(\ref{eqn12abx}) again, we see that $A_\mu$ satisfies the Lorentz condition,
\begin{equation}
\partial A_\mu(x) / \partial x_\mu = 0
\label{eqn14}
\end{equation}

\subsection{Number Operator}

\subsubsection{ Elementary Photon Theory }

The numbers operator for an elementary photon is defined as,
\begin{equation}
N_{\lambda}( {\bf p}) = b^\dagger_{\lambda}({\bf p}) b_{\lambda}({\bf p}).
\label{eqn12yx}
\end{equation}
When acting on a number state or Fock state, it returns the number of photons with momentum ${\bf p}$ and polarization $\lambda$.
\begin{eqnarray}
N_{\lambda}({\bf p}) (b^\dagger_{\lambda}({\bf p}))^m|0\rangle \;
= m (b^\dagger_{\lambda}({\bf p}))^m|0\rangle,
\label{eqn22abc}
\end{eqnarray}
for a state with $m$ photons. 
Normalizing in the
usual manner~\cite{koltun1},
\begin{eqnarray}
 b^\dagger_{\lambda}({\bf p})|n^{\lambda}_{\bf p} \rangle \;
= \sqrt{ (n^{\lambda}_{\bf p} +1)} 
|n^{\lambda}_{\bf p} +1\rangle,  \nonumber \\
b_{\lambda}({\bf p})|n^{\lambda}_{\bf p} \rangle \;
= \sqrt{ n^{\lambda}_{\bf p}} 
|n^{\lambda}_{\bf p} -1\rangle.
\label{eqn23qab}  
\end{eqnarray}
Acting on the one and zero particle states results in,
\begin{eqnarray}
b^\dagger_{\lambda}({\bf p})|0 \rangle \; = \: | 1^{\lambda}_{\bf p}\rangle, \nonumber \\
b_{\lambda}({\bf p})|1^{\lambda}_{\bf p}\rangle \; = \: |0\rangle. 
\label{eqn24aab}
\end{eqnarray}

\subsubsection{ Composite Photon Theory }

The number operators for right-handed and left-handed composite photons are defined as,
\begin{eqnarray}
N_R( {\bf p}) = G^\dagger_R({\bf p}) G_R({\bf p}), \nonumber \\
N_L( {\bf p}) = G^\dagger_L({\bf p}) G_L({\bf p}).
\label{eqn12xyv}
\end{eqnarray}       
Perkins~\cite{perkins1} showed that the effect of the composite
photon's number operator acting on a state of $m$ right-handed composite photons is,
\begin{equation}
N_R({\bf p}) (G^\dagger_R({\bf p}))^m|0\rangle \;
= \left( m - {m(m-1) \over \Omega }
\right) (G^\dagger_R({\bf p}))^m|0\rangle,
\label{eqn22ab}
\end{equation}
where $\Omega$ is a constant equal
to the number of states used to construct the wave packet,
and $N_R({\bf p})|0\rangle \; =  \:0$. 
This result differs from that for the elementary photon because of the second term, which is small for large $\Omega$. 
Normalizing, 
\begin{eqnarray}
 G^\dagger_R({\bf p})|n^R_{\bf p} \rangle \;
= \sqrt{ (n^R_{\bf p} +1) 
\left( 1- {n^R_{\bf p} \over \Omega} \right) }
|n^R_{\bf p} +1\rangle,  \nonumber \\
G_R({\bf p})|n^R_{\bf p} \rangle \;
= \sqrt{ n^R_{\bf p} 
\left( 1- {(n^R_{\bf p}-1) \over \Omega} \right) }
|n^R_{\bf p} -1\rangle,
\label{eqn23q}  
\end{eqnarray}
where $|n^R_{\bf p}\rangle$ is the state of $n^R_{\bf p}$ right-handed
composite photons having momentum ${\bf p}$ which is created
by applying $G^\dagger_R({\bf p})$  on the vacuum $n^R_{\bf p}$ times. 
Note that,
\begin{eqnarray}
G^\dagger_R({\bf p})|0 \rangle \; = \: | 1^R_{\bf p}\rangle, \nonumber \\
G_R({\bf p})|1^R_{\bf p}\rangle \; = \: |0\rangle, 
\label{eqn24ab}
\end{eqnarray}
which is the same result as obtained 
with boson operators.  The formulas in Equation~(\ref{eqn23q}) 
are similar to those in Equation~(\ref{eqn23qab}) with correction factors 
that approach zero for large $\Omega$.

\subsection{Commutation relations for E and H}

\subsubsection{ Elementary Photon Theory }

The commutation relations for electric and magnetic fields 
in the elementary photon theory are~\cite{schiff1},
\begin{eqnarray}
\left[ E_i(x),E_j(y) \right] 
= \left( \delta_{ij}{\partial \over \partial x_0} 
{\partial \over \partial y_0}-{\partial \over \partial x_i}
{\partial \over \partial y_j}  \right) 
\left[ i D(x-y) \right],
\label{eqn29abx}
\end{eqnarray}
\begin{eqnarray}
\left[ H_i(x),H_j(y) \right] = \left[ E_i(x),E_j(y) \right], 
\label{eqn29aby}
\end{eqnarray}
and
\begin{eqnarray}
\left[ E_i(x),H_j(y) \right] 
= -{\partial \over \partial y_0} \sum_{k = 1}^3
\epsilon_{ijk}{\partial \over \partial x_k}
\left[ i D(x-y) \right].
\label{eqn29abz}
\end{eqnarray}

\subsubsection{ Composite Photon Theory }

With the composite photon theory, the commutation relations for $E$ and $H$ are similar to the ones for the elementary photon theory. However, the extra terms in composite commutation relations~(\ref{eqn17}) result in extra terms for the $E$ and $H$ commutation relations~\cite{perkins1}. With the extra terms the commutation relations do not vanish for space-like intervals, indicating that composite particles have a finite extent in space~\cite{perkins1}. 

\begin{eqnarray}
\left[ E_i(x),E_j(y) \right] 
= \left( \delta_{ij}{\partial \over \partial x_0} 
{\partial \over \partial y_0}-{\partial \over \partial x_i}
{\partial \over \partial y_j}  \right) \nonumber \\
\left\{ i D(x-y) -{i \over {16 \pi^3}} \int {d^3 p \; \omega^{-1}_p 
sin \left[ p \cdot (x-y) \right] \left( \Delta_R ({\bf p}, {\bf p}) 
+ \Delta_L ({\bf p}, {\bf p}) \right)} \right\} \nonumber \\
-{i \over {16 \pi^3}} {\partial \over \partial y_0} \sum_{k = 1}^3
\epsilon_{ijk}{\partial \over \partial x_k} \int {d^3 p \; \omega^{-1}_p 
cos \left[ p \cdot (x-y) \right] \left( \Delta_R ({\bf p}, {\bf p}) 
- \Delta_L ({\bf p}, {\bf p}) \right)},
\label{eqn24abx}
\end{eqnarray}
\begin{eqnarray}
\left[ H_i(x),H_j(y) \right] = \left[ E_i(x),E_j(y) \right], 
\label{eqn24aby}
\end{eqnarray}
and
\begin{eqnarray}
\left[ E_i(x),H_j(y) \right] 
= -{\partial \over \partial y_0} \sum_{k = 1}^3
\epsilon_{ijk}{\partial \over \partial x_k} \nonumber \\
\left\{ i D(x-y) -{i \over {16 \pi^3}} \int {d^3 p \; \omega^{-1}_p 
sin \left[ p \cdot (x-y) \right] \left( \Delta_R ({\bf p}, {\bf p}) 
+ \Delta_L ({\bf p}, {\bf p}) \right)} \right\} \nonumber \\
-{i \over {16 \pi^3}} \left( \delta_{ij}{\partial \over \partial x_0} 
{\partial \over \partial y_0}-{\partial \over \partial x_i}
{\partial \over \partial y_j}  \right) \nonumber \\
\int {d^3 p \; \omega^{-1}_p cos \left[ p \cdot (x-y) \right] 
\left( \Delta_R ({\bf p}, {\bf p}) - \Delta_L ({\bf p}, {\bf p}) \right)}.
\label{eqn24abz}
\end{eqnarray}

\subsection{Charge Conjugation and Parity}

\subsubsection{ Elementary Photon Theory }

The antiphoton is identical to the photon. Thus the electromagnetic field can at most change by a factor of $-1$ under charge conjugation.
Since the electromagnetic current, 
${\bf j}_{\mu}(x)$, changes sign under the operation of charge conjugation, 
\begin{eqnarray}
C {\bf j}_{\mu}(x) = -{\bf j}_{\mu}(x),
\end{eqnarray}
the electromagnetic field must transform as,
\begin{eqnarray}
C {\bf A}_{\mu}(x) = -{\bf A}_{\mu}(x),
\end{eqnarray}
in order to leave the product ${\bf j}_{\mu}(x) \cdot {\bf A}_{\mu}(x)$ in the Lagrangian invariant. For ${\bf A}_{\mu}(x)$ in the plane-wave representation 
Equation~(\ref{eqn8.4ab}) this means, 
\begin{eqnarray}
C b_R({\bf p}) = -b_R({\bf p}), \nonumber\\
C b_L({\bf p}) = -b_L({\bf p}).
\label{eqn11ch} 
\end{eqnarray}
Under the parity operator 
the vector potential transforms as,
\begin{eqnarray}
P {\bf A}_{\mu}({\bf x},t) = {\bf A}_{\mu}(-{\bf x},t).
\end{eqnarray}
This implies that the creation and annihilations operators change as,
\begin{eqnarray}
P b_R({\bf p}) = b_L({\bf -p}), \nonumber\\
P b_L({\bf p}) = b_R({\bf -p}). 
\label{eqn11par}
\end{eqnarray} 
Under the combined operation of CP,
\begin{eqnarray}
CP {\bf A}_{\mu}({\bf x}, t) = -{\bf A}_{\mu}(-{\bf x}, t).
\end{eqnarray}
In short-hand notation,
\begin{eqnarray}
C \gamma = -\gamma, \nonumber \\
P \gamma = \gamma, \nonumber \\
CP \gamma = -\gamma.
\end{eqnarray}

\subsubsection{ Composite Photon Theory }

Under C (charge conjugation) and P (parity), the neutrino annihilation operator transform as follows:
\begin{eqnarray}
C a_2({\bf k}) = c_1({\bf k}), 
C c_2({\bf k}) = a_1({\bf k}), \nonumber \\
C a_1({\bf k}) = c_2({\bf k}), 
C c_1({\bf k}) = a_2({\bf k}), \nonumber \\
P a_2({\bf k}) = a_1({\bf -k}), 
P c_2({\bf k}) = c_1({\bf -k}),\nonumber \\ 
P a_1({\bf k}) = a_2({\bf -k}),
P c_1({\bf k}) = c_2({\bf -k}). 
\end{eqnarray}
We construct the composite antiphoton field in a manner similar to that of the composite photon field,
\begin{eqnarray}
{\overline A}_\mu(x) =  \sum_{\bf p} {1 \over 2\sqrt{ V \omega_p}}\left\{ 
\left[{\overline G}_R({\bf p}) 
{\overline u}^{-1}_{-1}({\bf p})i \gamma_{\mu} 
u^{+1}_{+1}({\bf p})
+ {\overline G}_L({\bf p}) {\overline u}^{+1}_{+1}({\bf p})i \gamma_{\mu} 
u^{-1}_{-1}({\bf p}) \right]e^{i p x} \right. \nonumber \\
\left. + \left[{\overline G}_R^\dagger({\bf p}) 
{\overline u}^{+1}_{+1}({\bf p})i \gamma_{\mu} u^{-1}_{-1}({\bf p})
+ {\overline G}_L^\dagger({\bf p}) {\overline u}^{-1}_{-1}({\bf p}) i\gamma_{\mu} u^{+1}_{+1}({\bf p}) \right]e^{-i p x}  \right\},
\label{eqn18.5}
\end{eqnarray}
with the annihilation operators for left-circularly and right-circularly polarized antiphotons with momentum ${\bf p}$ given by,
\begin{eqnarray}
{\overline G}_L( {\bf p}) = {1 \over \sqrt{2}} \sum_{\bf k} F^\dagger( {\bf k}) 
c_1( {\bf p} + {\bf k}) a_1({\bf -k})  \nonumber \\
{\overline G}_R( {\bf p}) = {1 \over \sqrt{2}} \sum_{\bf k} F^\dagger( {\bf k}) 
c_1({\bf -k}) a_1( {\bf p} + {\bf k}),
\label{eqn19}   
\end{eqnarray}
Note that ${\overline A}_{\mu}(x)$ contains the other two spinors from Equation~(\ref{eqn6}).
Appying the charge conjugation and parity operators on the composite photon annihilation operators gives,
\begin{eqnarray}
C G_L( {\bf p}) =-{\overline G}_L({\bf p}), \nonumber \\
C G_R( {\bf p}) =-{\overline G}_R({\bf p}), \nonumber\\  
P G_L( {\bf p}) = {\overline G}_R({\bf -p}), \nonumber \\
P G_R( {\bf p}) = {\overline G}_L({\bf -p}),   
\end{eqnarray}
where we have taken $F^{\dagger}({\bf k})$ to be symmetric in ${\bf k}$.
Applying the charge conjugation and parity operators on the composite photon field gives,
\begin{eqnarray}
C {\bf A}_{\mu}(x) = -{\bf {\overline A}}_{\mu}(x), \nonumber\\
P {\bf A}_{\mu}({\bf x},t) = {\bf {\overline A}}_{\mu}(-{\bf x},t),
\end{eqnarray}
since 
\begin{eqnarray}
{\overline u}^{+1}_{-1}({\bf p})i\gamma_{\mu} u^{-1}_{+1}({\bf p}) 
=-{\overline u}^{-1}_{-1}({\bf p})
i\gamma_{\mu} u^{+1}_{+1}({\bf p}),  \nonumber \\
{\overline u}^{-1}_{+1}({\bf p})i\gamma_{\mu} u^{+1}_{-1}({\bf p}) 
=-{\overline u}^{+1}_{+1}({\bf p})
i\gamma_{\mu} u^{-1}_{-1}({\bf p}), \nonumber \\
{\overline u}^{+1}_{-1}({\bf -p})i\gamma_{\mu} u^{-1}_{+1}({\bf -p}) = {\overline u}^{-1}_{+1}({\bf p})
i\gamma_{\mu} u^{+1}_{-1}({\bf p}).  
\end{eqnarray}
Under the combined operation of CP, 
\begin{eqnarray}
CP {\bf A}_{\mu}({\bf x}, t) = -{\bf A}_{\mu}(-{\bf x}, t).
\end{eqnarray}
In short-hand notation,
\begin{eqnarray}
C \nu_{2e} = \overline{\nu}_{1e}, \nonumber \\
C \overline{\nu}_{2e} = \nu_{1e}.
\end{eqnarray}
Since the internal structure of the composite photon is,
\begin{eqnarray}
\gamma = \nu_{2e} \overline{\nu}_{2e},
\end{eqnarray}
the antiphoton is, 
\begin{eqnarray}
\overline{\gamma} = \nu_{1e} \overline{\nu}_{1e}.
\end{eqnarray}
Not only is $\overline{\gamma}$ different than $\gamma$, but its neutrinos types have never been observed. Under C and P,
\begin{eqnarray}
C \gamma = -{\overline \gamma}, \nonumber \\
P \gamma = {\overline \gamma}. \nonumber \\
C {\overline \gamma} = -\gamma, \nonumber \\
P {\overline \gamma} = \gamma. 
\end{eqnarray}
The photon and antiphoton are invariant only under the combined operation of charge conjugation and parity,
\begin{eqnarray}
CP \gamma = \overline{\nu}_{2e} \nu_{2e} = -\gamma, \nonumber \\
CP \overline{\gamma} = \overline{\nu}_{1e}\nu_{1e} = -\overline{\gamma}.
\end{eqnarray}
However, there can be photon states that are eigenstates of C and P.
As is done with the neutral kaon, we create superpositions of the particle and antiparticle,
\begin{eqnarray}
|\gamma_1> = {1 \over \sqrt{2}} (|\gamma> + |\overline{\gamma}>) \nonumber \\
|\gamma_2> = {1 \over \sqrt{2}} (|\gamma> - |\overline{\gamma}>),
\label{eqn20}
\end{eqnarray}
Under charge conjugation,
\begin{eqnarray}
C |\gamma_1> = -|\gamma_1>, \nonumber \\
C |\gamma_2> = |\gamma_2>, 
\label{eqn21}
\end{eqnarray}
showing that $|\gamma_1>$ is an eigenstate of C with value -1,
while $|\gamma_2>$ is an eigenstate of C with value +1 with similar results under parity.
In the composite photon theory the electromagnetic field transforms in the usual way only under the combined operation of CP.

\subsection{Symmetry under Interchange}

\subsubsection{ Elementary Photon Theory }

Since the photon is its own antiparticle, all photons are identical. Thus, a state of two photons must be symmetric under interchange. This result has been used to rule out certain reactions~\cite{landau1,yang1}.

\subsubsection{ Composite Photon Theory }

In the composite theory, four photon states exist, i.e.,
$\gamma$, $\overline \gamma$, $\gamma_1$, and $\gamma_2$. If the photons are not identical, a state of two photons can be antisymmetric (as well as symmetric) under interchange. Therefore, a vector particle {\it can} decay into two photons~\cite{perkins4}. 

\subsection{\label{sec.photon_inter}Photon-Electron Interaction}

Here we examine Compton scattering, using Feynman diagrams. (The photo-electric effect is similar.) Fig.~1a shows the usual Feynman diagram for Compton scattering with the incoming photon imparting energy and momentum to an electron. Fig.~1b shows the same process with the photon replaced by the bound state of the neutrino-antineutrino pair as a chain of constituent fermion-antifermion bubbles. The local interaction is similar to that in Fermi's beta decay theory~\cite{wilson1}. The relevant Feynman rules are:

Incoming electron: $V^{-1/2} \Psi^{\lambda_1}_e({\bf p_1}), \lambda_1 = 1,2$

Outgoing electron: $V^{-1/2} {\overline \Psi}^{\lambda_2}_e({\bf p_2}), \lambda_2 = 1,2$

Propagator: $(-i \gamma_{\mu} p_{\mu} + m_e) / 
(p^2 + m_e^2)$

Incoming neutrino: $V^{-1/2} u^{+1}_{-1} ({\bf k_1})$

Incoming antineutrino: $V^{-1/2} {\overline u}^{-1}_{+1} ({\bf r_1})$

Outgoing neutrino: $V^{-1/2} {\overline u}^{+1}_{-1}({\bf k_2})$

Outgoing antineutrino: $V^{-1/2} u^{-1}_{+1} ({\bf r_2})$

Incoming photon: ${1 \over {\sqrt{2V \omega_k}}} \epsilon_{\mu}^{i}({\bf k_1})$

Outgoing photon: ${1 \over {\sqrt{2V \omega_k}}} \epsilon_{\mu}^{i*}({\bf k_2})$

Vertex: $-i e \gamma_\mu$

\begin{figure}
\includegraphics[scale=0.4]{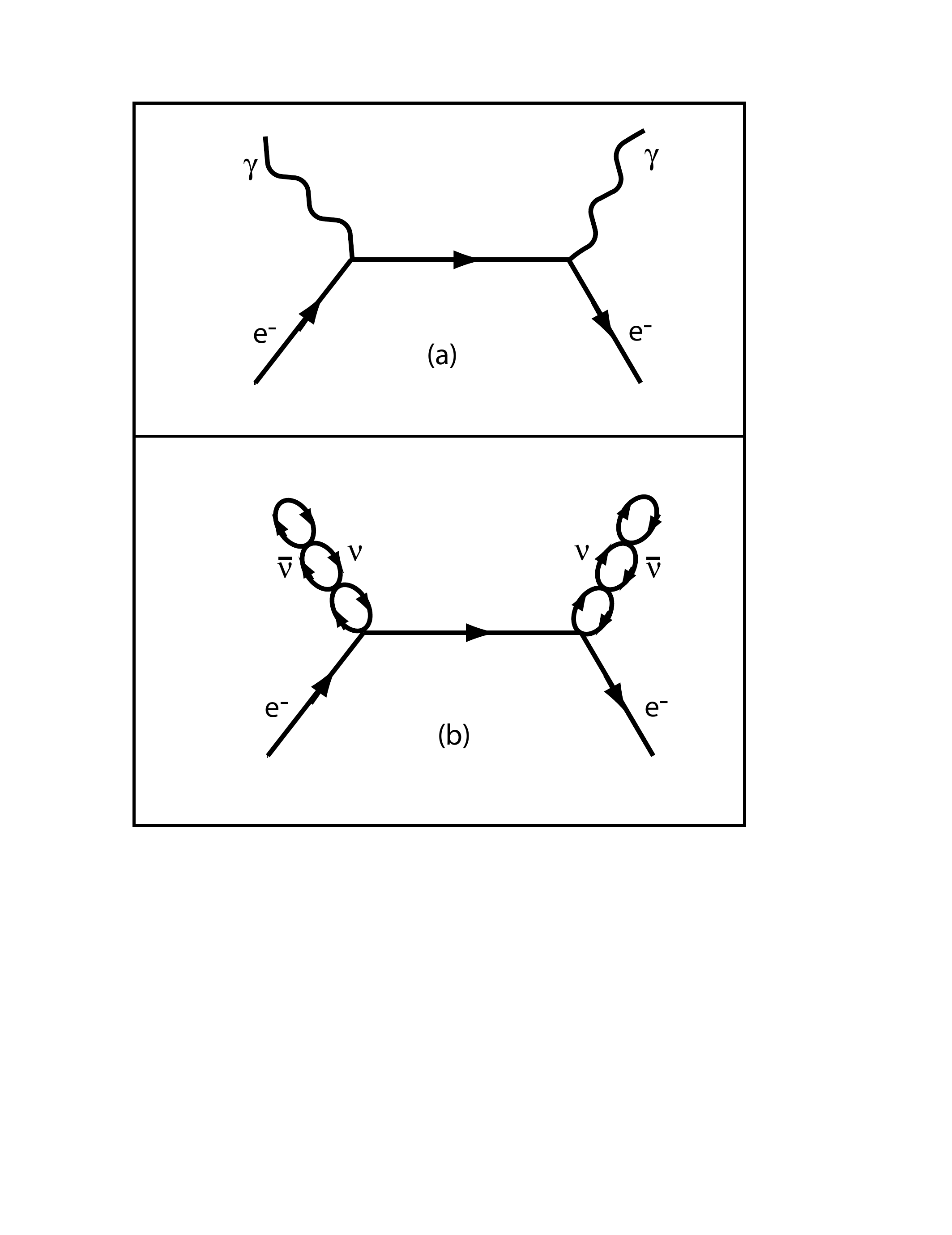}
\caption{ Compton scattering. (a) Elementary Photon Theory. (b) Composite Photon Theory.} 
\label{f1}
\end{figure}

\subsubsection{ Elementary Photon Theory }

The matrix element for Compton scattering as shown in Figure~\ref{f1} (a) is,
\begin{eqnarray}
{\cal M}= {-i e \over {2 V^2 \omega_p}}\sum_{{\bf q},\lambda_1,\lambda_2}
 \left\{ {\overline \Psi}^{\lambda_2}_e({\bf p_2}) \gamma_\mu \epsilon_{\mu}^{i*}({\bf k_2})
{{-i \gamma_{\mu} q_{\mu} + m_e} \over {(q^2 + m_e^2)}} \gamma_\mu
\epsilon_{\mu}^{i}({\bf k_1})\Psi^{\lambda_1}_e({\bf p_1}) \right\}.
\label{eqn24abzy}
\end{eqnarray}

\subsubsection{ Composite Photon Theory }

In the composite theory the matrix element for Compton scattering as shown in Figure~\ref{f1} (b) is,
\begin{eqnarray}
{\cal M}= {-i e \over {2 V^2 \omega_p}}
\sum_{{\bf q},\lambda_1,\lambda_2} \left\{ 
{\overline \Psi}^{\lambda_2}_e({\bf p_2}) \gamma_\mu u^{-1}_{+1}({\bf r_2})
{{-i \gamma_{\mu} q_{\mu} + m_e} \over {(q^2 + m_e^2)}}
{\overline u}^{+1}_{-1} ({\bf k_2}) \gamma_\mu \Psi^{\lambda_1}_e({\bf p_1}) 
{\overline u}^{-1}_{+1} ({\bf r_1}) \gamma_\mu u^{+1}_{-1}({\bf k_1}) \right. \nonumber\\  \left.
+ {\overline \Psi}^{\lambda_2}_e({\bf p_2}) \gamma_\mu u^{+1}_{-1}({\bf k_1})
{{-i \gamma_{\mu} q_{\mu} + m_e} \over {(q^2 + m_e^2)}}
{\overline u}^{-1}_{+1} ({\bf r_1}) \gamma_\mu \Psi^{\lambda_1}_e({\bf p_1}) 
{\overline u}^{+1}_{-1} ({\bf k_2}) \gamma_\mu u^{-1}_{+1}({\bf r_2})
 \right\}.
\label{eqn24bc}
\end{eqnarray}
The matrix element contains components,
\begin{eqnarray}
[{\overline \Psi}^{\lambda_2}_e({\bf p_2}) \gamma_\mu u^{-1}_{+1}({\bf r_2})]
[{\overline u}^{+1}_{-1} ({\bf k_2}) \gamma_\mu \Psi^{\lambda_1}_e({\bf p_1})], 
\label{eqn24cd}
\end{eqnarray}
and
\begin{eqnarray}
[{\overline \Psi}^{\lambda_2}_e({\bf p_2})\gamma_\mu u^{+1}_{-1}({\bf k_1})]
[{\overline u}^{-1}_{+1}({\bf r_1})\gamma_\mu \Psi^{\lambda_1}_e({\bf p_1})] 
\label{eqn24cde}
\end{eqnarray}
Since the electron-neutrino interaction is V-A, we must insert the projection operator, ${1\over{2}}(1 - \gamma_5)$ to select states with negative-helicity particles and positive-helicity antiparticles. With this insertion we have components,
\begin{eqnarray}
{1 \over 4}[{\overline \Psi}^{\lambda_2}_e({\bf p_2}) 
\gamma_\mu (1 - \gamma_5) u^{-1}_{+1}({\bf r_2})]
[{\overline u}^{+1}_{-1} ({\bf k_2}) \gamma_\mu (1 - \gamma_5) \Psi^{\lambda_1}_e({\bf p_1})], 
\label{eqn28}
\end{eqnarray}
and 
\begin{eqnarray}
{1 \over 4} [{\overline \Psi}^{\lambda_2}_e({\bf p_2})
\gamma_\mu (1 - \gamma_5)  u^{+1}_{-1}({\bf k_1})]
[{\overline u}^{-1}_{+1}({\bf r_1}) \gamma_\mu (1 - \gamma_5) \Psi^{\lambda_1}_e({\bf p_1})] 
\label{eqn29}
\end{eqnarray}
Since $u^{-1}_{+1}({\bf p})$ designates a positive-helicity antiparticle and
$ u^{+1}_{-1}({\bf p})$ designates a negative-helicity particle the insertion of ${1\over{2}}(1 - \gamma_5)$ does not change the result~\cite{perkins4}. 
However, for the interaction of an antiphoton with an electron,  
the terms contain components,
\begin{eqnarray}
{1 \over 4}[{\overline \Psi}^{\lambda_2}_e({\bf p_2}) 
\gamma_\mu (1 - \gamma_5) u^{-1}_{-1}({\bf r_2})]
[{\overline u}^{+1}_{+1} ({\bf k_2}) \gamma_\mu (1 - \gamma_5) \Psi^{\lambda_1}_e({\bf p_1})], 
\label{eqn28ab}
\end{eqnarray}
and 
\begin{eqnarray}
{1 \over 4} [{\overline \Psi}^{\lambda_2}_e({\bf p_2})
\gamma_\mu (1 - \gamma_5)  u^{+1}_{+1}({\bf k_1})]
[{\overline u}^{-1}_{-1}({\bf r_1}) \gamma_\mu (1 - \gamma_5) \Psi^{\lambda_1}_e({\bf p_1})] 
\label{eqn29ab}
\end{eqnarray}
The $(1 - \gamma_5) u^{+1}_{+1}({\bf p})$ and
$(1 - \gamma_5) u^{-1}_{-1}({\bf p})$ terms equate to zero as,
\begin{eqnarray}
{1 \over 2}(1 - \gamma_5) u^{+1}_{+1}({\bf p})
= \left( \begin{array}{cccc}
0 & 0 & 0 & 0 \\
0 & 0 & 0 & 0 \\ 
0 & 0 & 1 & 0 \\
0 & 0 & 0 & 1
\end{array} \right)
\sqrt{ {E + p_3} \over 2 E} 
\left( \begin{array}{c}
1 \\ {{p_1 + i p_2} \over {E + p_3}} \\
0 \\ 0 \\
\end{array} \right)
= \sqrt{ {E + p_3} \over 2 E} 
\left( \begin{array}{c}
0 \\ 0 \\
0 \\ 0 \\
\end{array} \right)
\label{eqnxyz} 
\end{eqnarray}
This indicates that antiphotons do NOT interact with elections in a matter world, because $\nu_{1e}$ and ${\overline \nu}_{1e}$ have the wrong helicity.

In an antimatter world, the positron-neutrino interaction is V+A and ${1\over{2}}(1 + \gamma_5)$ selects states with positive-helicity particles and negative-helicity antiparticles. In a symmetric manner
photons do not interact with positrons in an antimatter world~\cite{perkins4}. 

Experiment~\cite{badertscher} shows that all the photons in positronium are detected. Therefore, the photons involved must be $\gamma_1$ and $\gamma_2$, the superposition of $\gamma$ and $\overline{\gamma}$. 

Positrons interact with the electromagnetic field in a manner similar to that of electrons. Thus, the composite photon theory requires that the effect of virtual photons is the same in matter and antimatter worlds.

\section{\label{sec.conclusions}Conclusions}

In comparing the elementary and composite photon theories, it was noted that in the elementary theory it is difficult to describe the electromagnetic field with the four-component vector potential. This is because the photon has only two polarization states. This problem does not exist with the composite photon theory. The commutation relations are more complex in the composite theory because of the composite photon's internal fermion structure. However, this complexity is not unique to the composite photon; other composite particles with internal fermions have similar complexity. In the elementary theory the polarization vectors are chosen to give a transverse field, while in the composite theory they are determined by the fermion bispinors. The composite theory predicts Maxwell equations, while the elementary theory has been created to encompass it. Some differences are so slight that they are almost impossible to detect experimentally (i.e., Planck's law). However, the composite theory predicts that the antiphoton is different than the photon. 

Pryce~\cite{pryce1} had many arguments against a composite photon theory. His arguments are either not valid or irrelevant. Let us look at them one by one: 
1) Pryce: ``In so far as the failure of the theory can be traced to any one cause it is fair to say that it lies in the fact that light waves are polarized transversely while neutrino `waves' are polarized longitudinally.'' Both Case~\cite{case1} 
and Berezinski~\cite{berezinskii1} asserted that constructing transversely polarized photons is not a problem. The fact that one can combine neutrino fields and obtain a composite photon that satisfies Maxwell equations (as in Section~\ref{sec.maxwell}) proves that this is not a problem.
2) Pryce: ``In order to fix the representation, therefore, we must decide on a definite ${\bf a}$ [polarization vector perpendicular to ${\bf n}$]. This choice is entirely arbitrary, for among all unit vectors perpendicular to a given direction in space all are equivalent and none is singled out in any way.'' The composite theory singled out the two polarization vectors of Equation~(\ref{eqn12.5}) which are functions of ${\bf n}$. Under a rotation by an angle $\theta$ about ${\bf n}$ they change into themselves.
\begin{eqnarray}
{\bf \epsilon}^1_{\mu}( n) \rightarrow 
e^{i \theta}{\bf \epsilon}^1_{\mu}( n) \nonumber\\
{\bf \epsilon}^2_{\mu}( n) \rightarrow 
e^{-i \theta}{\bf \epsilon}^2_{\mu}( n)
\end{eqnarray} 
Note that ${\bf \epsilon}^1( n)$ is a self-orthogonal complex unit vector~\cite{ravndal1}.
3) Pryce: ``the theory [must] be invariant under a change of co-ordinate system $...$ it has been necessary to analyze rather carefully the transformation of the amplitudes under certain types of rotation and this reveals an  arbitrariness in the choice of certain phases.'' In order to obtain the completeness relation, Equation~(\ref{eqn15}), 
Kronig~\cite{kronig1} arbitrarily wrote his Equation (17) connecting neutrino spinors. Pryce showed that Kronig's Equation (17) combined with Kronig's Equation (19) is not invariant under a rotation of the coordinate system. Kronig's Equation (17) is not needed, as one can obtain the completeness relation, Equation~(\ref{eqn15}), from the plane-wave spinors as shown in Section~\ref{sec.composite_polar_vectors}.
Pryce's argument that the composite photon theory is not invariant under a rotation of coordinate system, applies to one unnecessary equation in Kronig's paper. 
4) Pryce: ``The conditions under which this will lead to a satisfactory theory of light are (1) that certain [Bose] commutation rules be satisfied; (2) that the theory be invariant under a change of coordinate system.'' Pryce required that composite photons satisfy Bose commutation relations. (Jordan and Kronig were working on that assumption.) 
Pryce~\cite{pryce1} showed that requiring $\left[\xi({\bf p}),\eta^\dagger({\bf q})\right] = 0$ meant that $\xi = 0$. For a proof using the last of Equation~(\ref{eqn17ba}), see~\cite{perkins2}. This is a valid point, but it is really irrelevant. Integral spin particles are considered to be bosons, and most integral spin particles (deuterons, helium nuclei, Cooper pairs, pions, kaons, etc.) are composite particles formed of fermions. These composite particles cannot satisfy Bose commutation relations because of their internal fermion structure, but their difference from perfect bosons is so small that it has not been detected, with the exception of Cooper pairs~\cite{lipkin1}. In the asymptotic limit, which usually applies, these composite particles are bosons.

An important test of these ideas will occur when the photons from anti-Hydrogen are examined. The composite photon theory predicts that the antiphotons from anti-Hydrogen will have the wrong helicity for interaction with electrons, and thus the antiphotons will not be detectable. Furthermore, ordinary photons have the wrong helicity for interaction with anti-Hydrogen.

\section{\label{sec.acknow} Acknowledgments}

Helpful discussions with Prof.~J.~E.~Kiskis are gratefully acknowledged.

\end{document}